\documentclass[12pt]{article}
\usepackage[dvips]{graphicx}
{\end{list}}
\newcounter{enumct}
\newenvironment{Enumerate}{\begin{list}{\arabic{enumct}.}%
{\usecounter{enumct}\setlength{\topsep}{0.2mm}%
\setlength{\partopsep}{0.2mm}\setlength{\itemsep}{0.2mm}%
\setlength{\parsep}{0.2mm}}}{\end{list}}

\begin{document}
\renewcommand{\thefootnote}{\fnsymbol{footnote}}
\begin{flushright}
KEK Preprint 98-208\\
NWU-HEP 98-02\\
TUAT-HEP 98-03\\
DPNU-98-48\\
TIT-HPE-98-013\\
OCU-HEP 98-01\\
PU-98-708\\
\end{flushright}
\begin{center}
{\bf \Large Measurement of the jet width in $\gamma \gamma$ collisions
            and \\ in $e^+ e^-$ annihilation process at TRISTAN
\footnote{To be published in Phys. Lett. {\bf B}.}
}
\vskip 0.5 cm
\vskip 0.5cm
\begin{center}
{\Large\bf TOPAZ Collaboration}
\end{center}

\vskip 3cm
\begin{abstract}
The shape of jets produced in (quasi-) real photon-photon
collisions as well as in $e^+ e^-$ annihilation process
has been studied with a cone jet finding algorithm,
using  the  data taken
with the TOPAZ detector at the TRISTAN $e^{+}e^{-}$ collider
at an average center-of-mass energy ($\sqrt{s_{ee}}$) of 58 GeV.
The results are presented in terms of the jet width as
a function of the jet transverse energy ($E_{T}^{jet}$) as well as a 
scaled transverse jet energy, $x_{T}$ (=2$\cdot
E_{T}^{jet}/\sqrt{s}$).
The jet width narrows as $E^{jet}_{T}$ increases; 
however, at the same value of $E_{T}^{jet}$ 
the jet width in $\gamma\gamma$ collisions at TRISTAN is significantly
narrower than that in $\gamma p $ collisions at HERA. 
By comparing our results with the data in other reactions,
it has been shown that the jet width in $\gamma \gamma$, $\gamma p$, 
$ p \bar{p} $ collisions as well as the $e^+ e^-$ annihilation process
has an approximate scaling behavior as a  function of $x_{T}$.
\end {abstract}

\newpage
\begin{center}
{\Large The TOPAZ Collaboration}
\end{center}

\vskip 0.5cm
K.Adachi$^{(1)}$,
H.~Hayashii$^{(1)}$\footnote{e-mail address:
hayashii@hepl.phys.nara-wu.ac.jp},
K.~Miyabayashi$^{(1)}$,
S.~Noguchi$^{(1)}$,
A.~Miyamoto$^{(2)}$,
T.~Tauchi$^{(2)}$,
K.~Abe$^{(3)}$,
T.~Abe$^{(3)}$\footnote{Present address: Stanford Linear Accelerator Center,
Stanford University, Stanford, California 94309,USA},
I.~Adachi$^{(2)}$,
M.~Aoki$^{(3)}$,
M.~Aoki$^{(4)}$,
R.~Enomoto$^{(2)}$,
K.Emi$^{(2)}$,
H.~Fujii$^{(2)}$,
K.~Fujii$^{(2)}$,
T.~Fujii$^{(6),(10)}$,
J.~Fujimoto$^{(2)}$,
N.~Fujiwara$^{(1)}$,
H.~Hirano$^{(5)}$,
B.~Howell$^{(8)}$,
N.~Iida$^{(2)}$,
H.~Ikeda$^{(2)}$,
Y.~Inoue$^{(7)}$,
S.~Itami$^{(3)}$,
R.~Itoh$^{(2)}$,
H.~Iwasaki$^{(2)}$,
M.~Iwasaki$^{(1)}$\footnote{Present address: University of Oregon, Eugene, OR 97403, USA},
R.~Kajikawa$^{(3)}$,
K.~Kaneyuki$^{(4)}$,
S.~Kato$^{(9)}$,
S.~Kawabata$^{(2)}$,
H.~Kichimi$^{(2)}$,
M.~Kobayashi$^{(2)}$,
D.~Koltick$^{(8)}$,
I.~Levine$^{(8)}$,
H.~Mamada$^{(5)}$,
K.~Muramatsu$^{(1)}$,
K.~Nagai$^{(10)}$\footnote{Present address: EP division, CERN, CH-1211,
Geneva 23, Switzerland},
K.~Nakabayashi$^{(3)}$,
M.~Nakamura$^{(7)}$,
E.~Nakano$^{(7)}$,
O.~Nitoh$^{(5)}$,
A.~Ochi$^{(4)}$,
F.~Ochiai$^{(11)}$,
N.~Ohishi$^{(3)}$,
Y.~Ohnishi$^{(2)}$,
Y.~Ohshima$^{(4)}$,
H.~Okuno$^{(2)}$,
T.~Okusawa$^{(7)}$,
E.~Shibata$^{(8)}$,
A.~Sugiyama$^{(3)}$,
H.~Sugiyama$^{(1)}$,
S.~Suzuki$^{(3)}$,
K.~Takahashi$^{(5)}$,
T.~Takahashi$^{(7)}$,
T.~Tanimori$^{(4)}$,
Y.~Teramoto$^{(7)}$,
M.~Tomoto$^{(3)}$,
T.~Tsukamoto$^{(2)}$,
T.~Tsumura$^{(5)}$,
S.~Uno$^{(2)}$,
Y.~Watanabe$^{(4)}$, 
A.~Yamamoto$^{(2)}$, and
M.~Yamauchi$^{(2)}$\\

\vskip 0.5cm
{\it
$^{(1)}$Department of Physics, Nara Women's University,
 Nara 630-8506, Japan
\\
$^{(2)}$KEK, High Energy Accelerator Research Organization,Tsukuba,
  Ibaraki 305-0801, Japan
\\
$^{(3)}$Department of Physics, Nagoya University,
 Nagoya 464-8601, Japan
\\
$^{(4)}$Department of Physics, Tokyo Institute of Technology,
     Tokyo 152-8551, Japan
\\
$^{(5)}$Dept. of Applied Physics,
Tokyo Univ. of Agriculture and Technology,
 Tokyo 184-8588, Japan
\\
$^{(6)}$Department of Physics, University of Tokyo,
   Tokyo 113-0033, Japan
\\
$^{(7)}$Department of Physics, Osaka City University,
 Osaka 558-8585, Japan
\\
$^{(8)}$Department of Physics, Purdue University,
 West Lafayette, IN 47907, USA
\\
$^{(9)}$Institute for Nuclear Study, University of Tokyo,
   Tokyo 188-0002, Japan
\\
$^{(10)}$The Graduate School of Science and Technology,
Kobe University,
Kobe 657-8501, Japan
\\
$^{(11)}$Faculty of Liberal Arts, Tezukayama Gakuin University,
 Nara 631, Japan
}
\end{center}

\newpage
\section{Introduction}
\indent
At $e^+e^-$ colliders copious photons are emitted
from the beam electrons or positrons, where most of the photons carry
only a 
small four-momentum squared $(Q^{2})$, and can be considered to be
quasi-real ($Q^{2}\approx 0$).
These quasi-real photons can interact with each other and produce
hadrons in the final state. 
Before TRISTAN experiments, hadron production in
quasi-real photon-photon ($\gamma\gamma$) collisions
had generally been thought to be mainly caused by 
soft processes, and that perturbative-QCD can not be applied.
This situations drastically changed when the production
of high-transverse-energy ($E_{T}^{jet}$) jets was observed
in $\gamma\gamma$ interactions at TRISTAN\cite{HAYASHII,AMY1} and
LEP\cite{ALEPH} $e^+e^-$ colliders.

\begin{figure}[ht]
\begin{center}
\includegraphics*[height=15cm,clip]{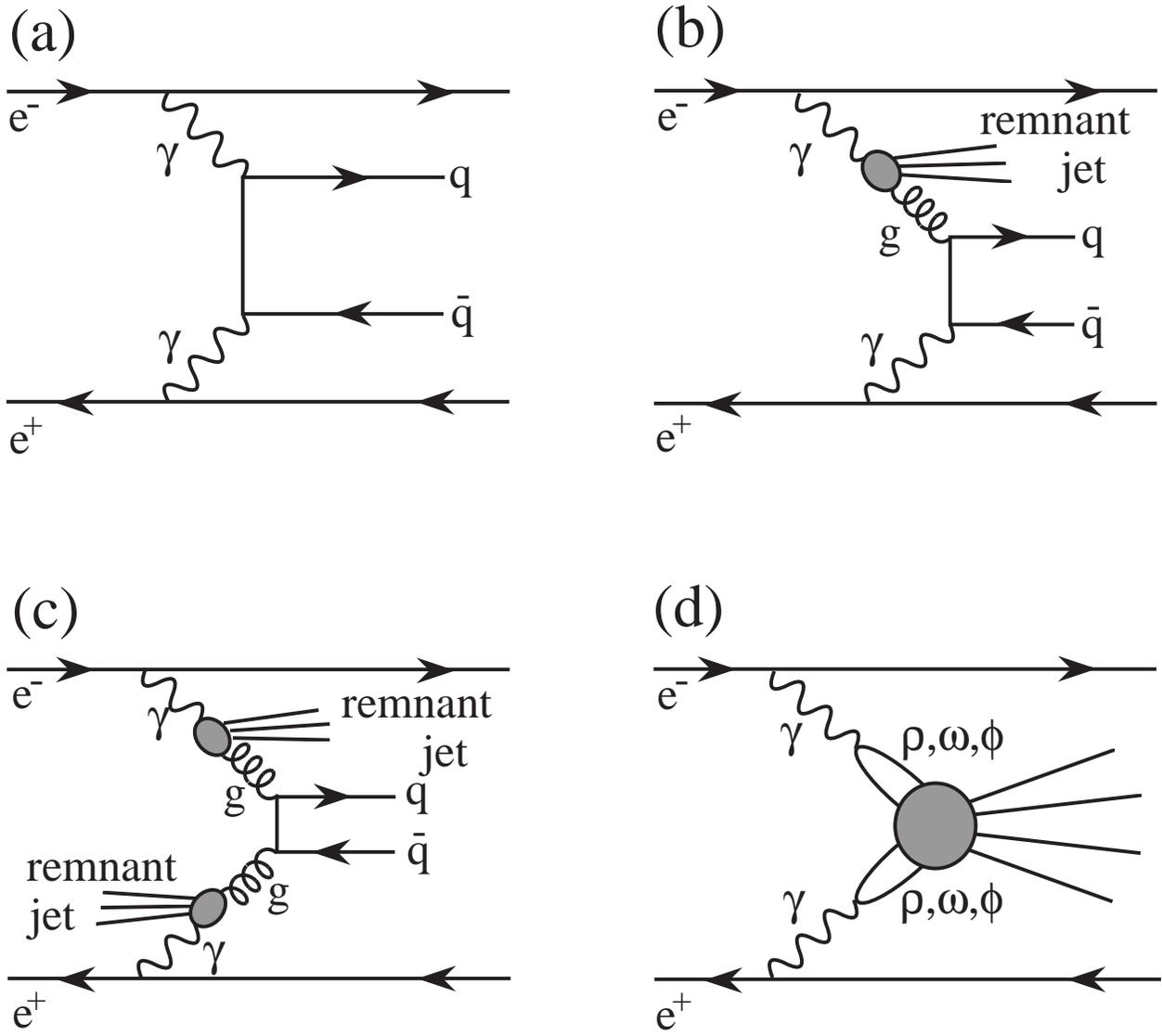}
\caption{
         Examples of diagrams contributing to the hadron production in
         $\gamma \gamma$ collisions: (a) direct(QPM) process, (b)
         one-resolved process, (c) two-resolved process and (d)
         soft VDM process.}
\label{fig:fig1}
\end{center}
\end{figure}

\vspace{5 mm}
High-$E_{T}^{jet}$ jet production in 
$\gamma\gamma$ collisions is an interesting field to study the
photon structure and to test perturbative QCD predictions.
Since the photon can fluctuate into hadronic components
before  interactions, three classes of hard QCD processes
are expected to contribute to jet production
at the leading order (LO)\cite{gamgam,x}:
either as a direct process where two photons directly interact 
(Fig.\ref{fig:fig1}a),
one-resolved process where a bare photon interacts with a
parton (quark or gluon) of the other photon
(Fig.\ref{fig:fig1}b), or a two-resolved
process where partons of both photons interact
(Fig.\ref{fig:fig1}c).
The jet cross sections as a function of  $E_{T}^{jet}$ were measured
up to 8 GeV
by TOPAZ\cite{HAYASHII} and AMY\cite{AMY1}, and were compared
with the next-leading-order QCD calculations in refs.\cite{aure94}
and \cite{kle96}. Recently, the jet cross sections were measured
up to 17 GeV at LEP2 by OPAL\cite{OPAL}.
The existence of the remnant-jet,  i.e. accompanied hadronic
activities
in the low-angle region, is a direct signature of the resolved
processes. This signature
has been observed in TOPAZ\cite{HAYASHII,photon98} and OPAL\cite{OPAL}
experiments. It has been reported that the charm-quark production
in $\gamma\gamma$ collisions requires the resolved-photon
processes\cite{charm}.
The necessity of gluon contents in the photon is emphasized in
ref.\cite{HAYASHII,charm}.
In photon-hadron interactions, similar QCD predictions have already
been confirmed at the HERA $ep$ collider\cite{ZEUS,H1}.

\vspace{5 mm}
In addition to production rates of jets, 
the internal structure of jets,
i.e. jet shape or jet profile,
is expected to provide additional  information about the QCD dynamics
\cite{eks92}-\cite{jetwidth3}. 
The shape might depend on the type of primary parton; quark or gluon.
However, it also depends on the kinematical variables, such as the 
transverse energy
and rapidity of the jets, as well as the jet algorithm.
Theoretically, the jet shape is related to  the ratio of
the higher-order cross sections divided by the leading-order cross
sections\cite{jetwidth1,jetwidth3}. Therefore, the next-leading-order 
(NLO) calculation of the jet rates
provides leading-order predictions on the jet shape\cite{jetwidth3}.
Although NLO predictions on the jet shape in hadron (or photon)
interactions are not yet available,
it is suggested in ref.\cite{jetwidth1} that
the jets in hadron-hadron, photon-hadron
and photon-photon collisions at different c.m. energies $(\sqrt{s})$ 
should have the same shape
for an equal value of the scaled variable
$x_{T}=\frac{2 E_{T}^{jet}}{\sqrt{s}}$ 
at the fixed value of the jet pseudorapidity($\eta^{jet}$)
\footnote{This expectation is based on the following
          arguments\cite{jetwidth1}.
          The $E_{T}^{jet}$ dependence in the jet shape 
          is mainly caused by the dependence in the hard sub-process
          cross sections. However, the  sub-process
          cross sections depend only on dimensionless variables,
          $x_{T}$, $\eta$ and the momentum fraction $x_{a,b}$ of the
          incoming partons and other dimensional factors cancel in the
          ratio.
          Since $x_{a,b}$ have been integrated over in the observable,
          the jet shape depends only on $x_{T}$  for fixed 
          $\eta^{jet}$.
          In the experiments, the jet shape in the central region
          ($|\eta^{jet}| < 1.0$)
           has been mainly measured so far.}.
A small deviation from the scaling behavior is
expected only through the $E_{T}^{jet}$ dependence in
the strong-coupling constant $\alpha_{s}(\mu) (\mu \sim E_T^{jet})$.

\vspace{5 mm}
In this paper, we study the $E_{T}^{jet}$ dependence of the
jet shape produced both in $\gamma\gamma$ collisions and in $e^+e^-$
annihilation process at the hadron level.
The cone jet finding algorithm is applied for both reactions.

In the jet cone algorithm, so far, two kinds of jet shape variables, 
i.e. the jet profile function\cite{eks92}
and the jet width, have been used in experiments.
The jet profile function is given by the fraction of the jet energy
within an inner cone of a certain size.
This has been used in $p\bar{p}$ collisions at
TEVATRON\cite{CDFppwidth},
$\gamma p$  collisions by the ZEUS collaboration at
HERA\cite{ZEUSgpwidth}
and $\gamma\gamma$\cite{OPAL} and $e^+e^-$\cite{OPALeewidth}
collisions by the OPAL collaboration at LEP.
On the other hand, the jet width was used for the first time in
studies of
$\gamma p$ collisions by the H1 collaboration at HERA\cite{H1gpwidth}.
They applied the same method to the  $p \bar{ p}$ data in UA1
and  observed a scaling behavior of the jet width
for the jets in $\gamma p$ and $p \bar{p} $ collisions\cite{H1gpwidth}.
 
\vspace{5 mm}
We here extend  the H1-method 
for the jets produced in $\gamma\gamma$
collisions and $e^+e^-$ annihilations. We measured the jet width
as a function of $E_{T}^{jet}$ for both inclusive-jet and dijet
samples. The width was measured up to $E_{T}^{jet}=8$ GeV 
for $\gamma\gamma$ collisions,
which corresponds to 0.3 in  $x_{T}=\frac{2\cdot
E^{jet}_{T}}{\sqrt{s_{ee}}}$.
The region of $x_{T} \sim 1 $ could be measured
by using $e^+ e^-$ annihilation events.
The jet width for direct and resolved samples in $\gamma\gamma$ 
collisions was also studied separately
by reconstructing the momentum-fraction of the parton
inside the photon for  dijet events.

\section{Experimental data }
\indent
The data used in this analysis were taken with the TOPAZ detector 
at the TRISTAN $e^{+}e^{-}$ collider (KEK) at
an average center-of-mass energy of 58.0 GeV.
All data taken for the period from 1990 to 1995 were used in
this analysis, which corresponds to
an integrated luminosity of 287.8 pb$^{-1}$.

\vspace{5 mm}
A detailed description of the TOPAZ detector can be found
elsewhere \cite{TOPAZ}-\cite{trig}.
In this analysis, 
both charged tracks and the neutral clusters were used for event
selection and jet reconstruction.
Charged tracks were measured in a 1.0 Tesla magnetic field 
with the time-projection-chamber (TPC)
in the angular region of  $ |\cos{\theta}| \leq$ 0.85,
where $\theta$ is the polar angle of a particle with respect to the
beam axis. TPC provided a momentum resolution of 
$\sigma_{p_t}/p_{t} = \sqrt{(1.5p_t)^2+1.6^2}$(\%),
where $p_{t}$ (in GeV) is the transverse momentum to the beam axis.  
Electromagnetic showers were detected with three kinds of
calorimeters: 
a barrel lead-glass calorimeter (BCL), an end-cap 
Pb-proportional-wire-counter-sandwich calorimeter (ECL) and
a forward bismuth-germanate-crystal 
calorimeter (FCL). Each detector covered polar angular ranges of 
$|\cos{\theta}| \leq 0.85$, 
$0.85 \leq | \cos{\theta} | \leq 0.98$ 
and $0.972 \leq | \cos{\theta} | \leq 0.998$ $(=3.2^{\circ})$,
respectively.
Since  FCL is located very close to the beam pipe, the calorimeter is
protected from beam-induced backgrounds by an extensive shielding 
system \cite{mask}.

In $\gamma \gamma$ collisions, the detection efficiencies are
sensitive to the  trigger condition of charged tracks because of their
low momentum and low multiplicity of tracks\cite{trig}.
Through the experiments, the conditions of the charged track-trigger 
were changed according to the beam conditions. There must be at least
two tracks with $p_t > $ 0.3 GeV $\sim$ 0.7 GeV; also 
the minimum opening angle of the
two highest $p_t$ tracks is 45$^{\circ}$ $\sim$ 70$^{\circ}$.
The change in the trigger conditions was taken into account
in the TOPAZ simulation program. The effects 
on the jet width studied in this analysis were found to be
negligible ($<1\%$).

\section{Event selection and jet reconstruction}
\indent
The hadronic events produced in $\gamma\gamma$ collisions are selected
in the following criteria:
\begin{Enumerate}
\item The number of charged tracks with $p_t > 0.1$ GeV and the polar
      angle $|\cos\theta| < 0.83 $ should be at least 4;
\item The position of the event vertex 
      should be within 3.0 cm 
      in the r$\phi$-plane and within $\pm$ 3.0 cm along the beam line 
      from the interaction point; 
\item The visible energy ($E_{vis}$) of the event 
      should satisfy  $E_{vis}\leq 35$ GeV, 
      where both the charged tracks in TPC and the neutral clusters 
      in BCL and ECL are
      used in the calculation of $E_{vis}$;
\item The mass of the system of the observed hadrons ($W_{vis}$) should
      be  $W_{vis}\geq 2$ GeV, where the tracks in TPC 
      and the clusters in BCL and ECL
      are used; and
\item The energy of the most energetic cluster in BCL,
      ECL, or FCL should be less than 0.25 $\times E_{beam}$. 
\end{Enumerate}
These selection criteria leave 286k multihadron events.
The criterion 5 ensures the anti-tag condition,
which limits the scattering angle of the beam electrons to be less
than $3.2^{\circ}$. With this anti-tag requirement,  
the virtuality ($Q^2$) of the 
photons ranges  from $\sim 10^{-8}{\rm \ GeV^{2}}$
to $2.6\  {\rm \ GeV}^{2}$. The mean value of $Q^{2}$
is $\sim 10^{-4}$ GeV$^2$.

For the hadronic events in $e^+e^-$ annihilation process,
the standard cuts which have been used in the
TOPAZ collaboration were applied to 278.0 pb$^{-1}$ of
data. The criteria used in the selection
are described in ref.\cite{hadsel}. 
A total of 29k $e^+e^-$ annihilation events were selected.

\vspace{5 mm}
In order to identify jets in the hadronic events both in
$\gamma\gamma$ collisions and $e^+e^-$ annihilation process,
we used the conventional jet cone algorithm\cite{snowmass}, 
where a jet is defined as a large amount of transverse
energy (or momentum) concentrated in a cone of radius $R$
in  the $\eta$(pseudorapidity
\footnote{Pseudorapidity $\eta$ is defined in terms of the
          polar angle $\theta$ as $\eta \equiv$ - ln tan($\theta$/2).
          The electron-beam direction is taken to be the + z direction
          in this experiment.}
) - $\phi$(azimuthal angle) plane. A particle $i$ is included in a jet
if it lies within the cone
\begin{equation}
\sqrt{(\eta^{jet}-\eta^{i})^2 + 
                 (\phi^{jet}-\phi^{i})^2} \leq R.
\end{equation}
In the cone algorithm,
the transverse energy of the jet ($E_T^{jet}$) is
calculated as a scalar sum of the particle transverse
energy ($E_{t_i} = E_i \cdot \sin\theta_i$) 
in a cone,
\begin{equation}
E_T^{jet}   =  \sum_{i\in {\rm cone}} E_{t_i},
\end{equation}
and the jet direction($\eta^{jet}$ and $\phi^{jet}$) is defined by the
following weighted averages:
\begin{equation}
\eta^{jet}  =  \frac{1}{\displaystyle  E_{T}^{jet}}   
                   \displaystyle \sum_{i\in {\rm cone}} E_{t_i} \cdot \eta_{i}
\quad {\rm and} \quad
\phi^{jet}  =  \frac{1}{\displaystyle  E_{T}^{jet}}
\displaystyle \sum_{i\in {\rm cone}}  E_{t_i} \cdot \phi_{i} .
\end{equation}
The jet direction is determined by  iteration. The iteration is stopped
when particles contained in the cone are not changed in the next
iteration.

When two or more jets overlap each other, particles in the overlapped
region were assigned to the highest $E_{T}^{jet}$ jet in this analysis
\footnote{This point is different from our previous
          analysis\cite{HAYASHII}, where two jets are combined if
          their directions satisfy the condition recommended by 
          Ellis {\it et al.} in ref.\cite{snowmass}. However, we
          do not use this method in this analysis, because when there
          are more than three seed-jets, they are sometimes
          recombined into one jet, and the size of the jet cone
          becomes very big by the iteration of the recombination.
          This mainly happens when 
          there are  activities from  remnant-jets in an event.}. 
In jet reconstruction, all charged particles with a transverse
momentum greater than 0.10 GeV and neutral particles with an energy
greater than 0.3 GeV were used.
In this analysis, the size of the cone radius($R$) was fixed at
$R=1.0$.

\vspace{5 mm}
The detector effects on the jet $E_{T}^{jet}$ and jet directions
$(\eta^{jet},\phi^{jet})$ were checked by using
Monte-Carlo samples generated by the PYTHIA program
(version: 5.720)\cite{PYTHIA}. The results show that the
jet directions$(\eta^{jet},\phi^{jet})$ have good correlations
between the generated and observed quantities.
The resolutions are
$\sigma_{\eta^{jet}}=0.05$ and $\sigma_{\phi^{jet}}= 0.06$ 
for $\eta^{jet}$ and $\phi^{jet}$, respectively. 
However, $E_{T}^{jet}$ in the observed level
is systematically lower
than the generated (hadron level) one by about $13\%$. This effect has
been corrected in the data.
After correcting $E_{T}^{jet}$,
39,829 inclusive-jets in $\gamma\gamma$ collisions
are selected with the conditions of
$E_T^{jet} \geq$ 2.0 GeV and $|\eta^{jet}| \leq$ 0.7.
Dijet events are selected by taking an event with two or more
jets with fulfilling that the highest $E_T^{jet}$ be greater than 3.0
GeV
and the second highest $E_T^{jet}$ be greater than 2.0 GeV\cite{Et12}.
The pseudorapidity of both jets is required to be within
$|\eta^{jet}| \leq$ 0.7.
With this condition, 3,582 dijet events are selected.

\vspace{5 mm}
The beam-gas background remaining in the inclusive-jet and dijet
samples was estimated to be 11.2\% and 0.70\%, respectively,
from the number of events in the side-band of the
event-vertex distribution along the beam-axis.
The physical background mainly comes from $e^+e^-$ annihilation
events:
$e^+e^- \to q\bar{q}(\gamma)$. The contribution of this background
has been estimated from the KORALZ\cite{KORALZ} and PYTHIA programs to
be 9.54\% and 18.4\% for inclusive-jet and dijet samples, respectively. 
The other physical backgrounds 
($e^+e^- \to \tau^+\tau^-(\gamma)$, 
$e^+e^- \to e^+e^-\tau^+\tau^-$ and $e^+e^- \to e^+e^-e^+e^-$) were
found to be 
0.19\% (0.61\%), 1.21\% (2.48\%) and 0.28\% (0.86\%) for the
inclusive-jet (dijet) samples.
Hereafter, these backgrounds are subtracted from data on a bin-by-bin
basis.

\section{ Direct and resolved contributions in dijet events}
\indent
We first demonstrate how the direct and resolved contributions
can be separated in the dijet sample in $\gamma\gamma$ collisions. 
This separation is a useful tool to study the  jet width
from direct and resolved processes independently.
Other detailed studies of the hadronic final state in this dijet
sample are given in ref.\cite{photon98}.

\vspace{5 mm}
In the leading-order QCD, two hard partons are produced
in $\gamma \gamma$ collisions.
In the one- or two-resolved processes, two high-$E_{T}^{jet}$
jets are expected to be accompanied by one or
two remnant jets in the near beam-direction\cite{gamgam}.
A pair of
variables, $x_{\gamma}^+$ and $x_{\gamma}^-$ can then be defined 
which specify the fraction of the photon momentum
carried by the parton inside the photon in
hard interactions.  One can calculate $x_\gamma^{\pm}$ 
 from the observables with a good approximation \cite{x}:
\begin{equation}
x^{+}_{\gamma} = \frac{\displaystyle \sum_{j\in {\rm jets}}(E - p_z)_j}
                     {\displaystyle \sum_{i\in {\rm hadrons}}(E - p_z)_i}
\quad {\rm and} \quad
  x^{-}_{\gamma} = \frac{\displaystyle \sum_{j\in {\rm jets}}(E + p_z)_j}
                     {\displaystyle \sum_{i\in {\rm hadrons}}(E + p_z)_i},
\label{eq:x}
\end{equation}
where $p_z$ is the momentum component along the
$z$-axis of the detector and $E$ is the energy of each particle or
jet. The numerator of Eq.(\ref{eq:x}) is the sum for particles within
the jet cone, while the denominator is the sum for all particles in a
dijet event.
Since the direction of the electron beam is taken to be the positive
direction of the $z$-axis in the TOPAZ convention,
$x^{-}$($x^{+}$) corresponds to the parton-momentum fraction
of the photon radiated from electron (positron) beam.
We introduce a variable,
$x^{min}_{\gamma}$, which is defined as the smaller value of 
$x^{-}_{\gamma}$ and $x^{+}_{\gamma}$.

\begin{figure}[ht]
\begin{center}
\includegraphics*[height=15cm,clip]{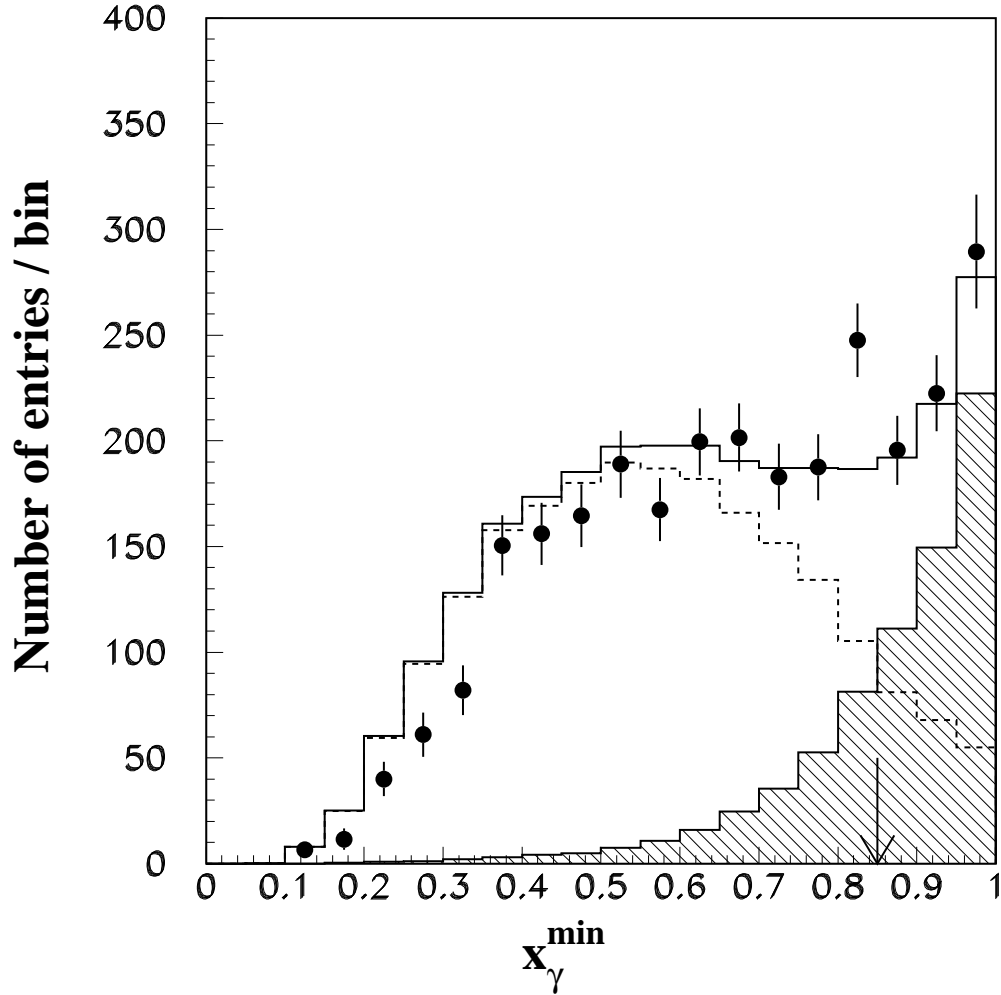}
\caption{
         $x_{\gamma}^{min}$ distribution for dijet events
         in $\gamma\gamma$ collisions. 
         Data (solid circles) are compared to the predictions
         of PYTHIA, where the GRV-LO
         photon structure function with $p_{t}^{min}$=1.6 GeV is used.
         The hatched histogram is the
         contribution from the direct process, and the dashed one
         is that from the resolved processes. The solid one is the
         sum of both processes. The arrow 
         at $x^{min}_{\gamma}$ = 0.85 indicates the division
         between direct($>0.85$) and resolved($<0.85$) events.
         Only statistical errors are shown.
         The Monte-Carlo predictions are normalized to the data
         luminosity by taking into account the predicted cross
         sections.}
\label{fig:x}
\end{center}
\end{figure}

\vspace{5 mm}
In Fig.\ref{fig:x}, the distribution of $x^{min}_{\gamma}$
is compared with
the Monte-Carlo predictions of the PYTHIA (5.720) program,
where no correction for the selection cut and the detector effects
has been applied, but the background has been subtracted.
The contribution
from the direct and resolved processes, predicted by PYTHIA, are also
shown separately. 
The predictions of PYTHIA are normalized to the luminosity of data 
by taking into account the predicted cross sections. 
The clear peak seen in the vicinity of 
$x^{min}_{\gamma} \sim$ 1 can be well-explained from the
direct component.
In the low-$x_{\gamma}^{min}$ region 
($ x^{min}_{\gamma} \leq$ 0.8), where the contribution from the
resolved
processes dominates, it is found that the overall shape as well as its
magnitude are well explained by  PYTHIA with 
the GRV-LO photon structure function\cite{GRV},
though for the region around 0.2 $\leq x^{min}_{\gamma} \leq$ 0.6 the
prediction is about 20\% higher than the data.
(The comparison with other cases are given in ref.\cite{photon98}).
%

In an analysis of the jet width, the direct and resolved
samples are separated at $x^{min}_{\gamma}$ = 0.85.
In the Monte-Carlo simulation of PYTHIA, about 67\% of the 
events from the direct process are contained in the 
$x^{min}_{\gamma} >$ 0.85 
region, while about 90\% of the events from the resolved process are
contained in the $x^{min}_{\gamma} <$ 0.85 region.

\section{Transverse energy flow and jet width}
\begin{figure}[ht]
\begin{flushleft}
\begin{minipage}[htbp]{7.5 cm}
\includegraphics*[height=8.0cm]{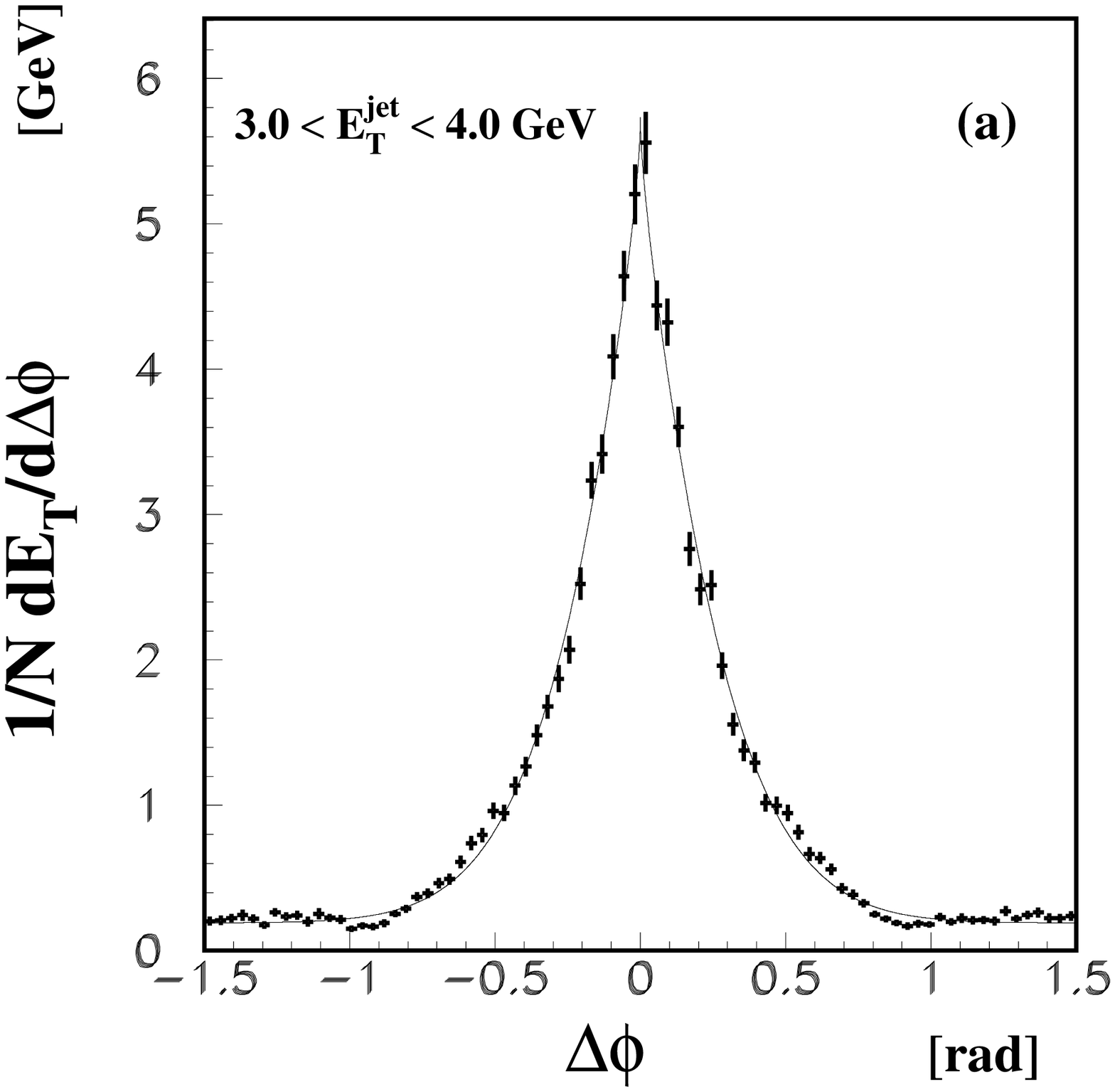}
\end{minipage}
\quad
\begin{minipage}[htbp]{7.5 cm}
\includegraphics*[height=8.0cm ]{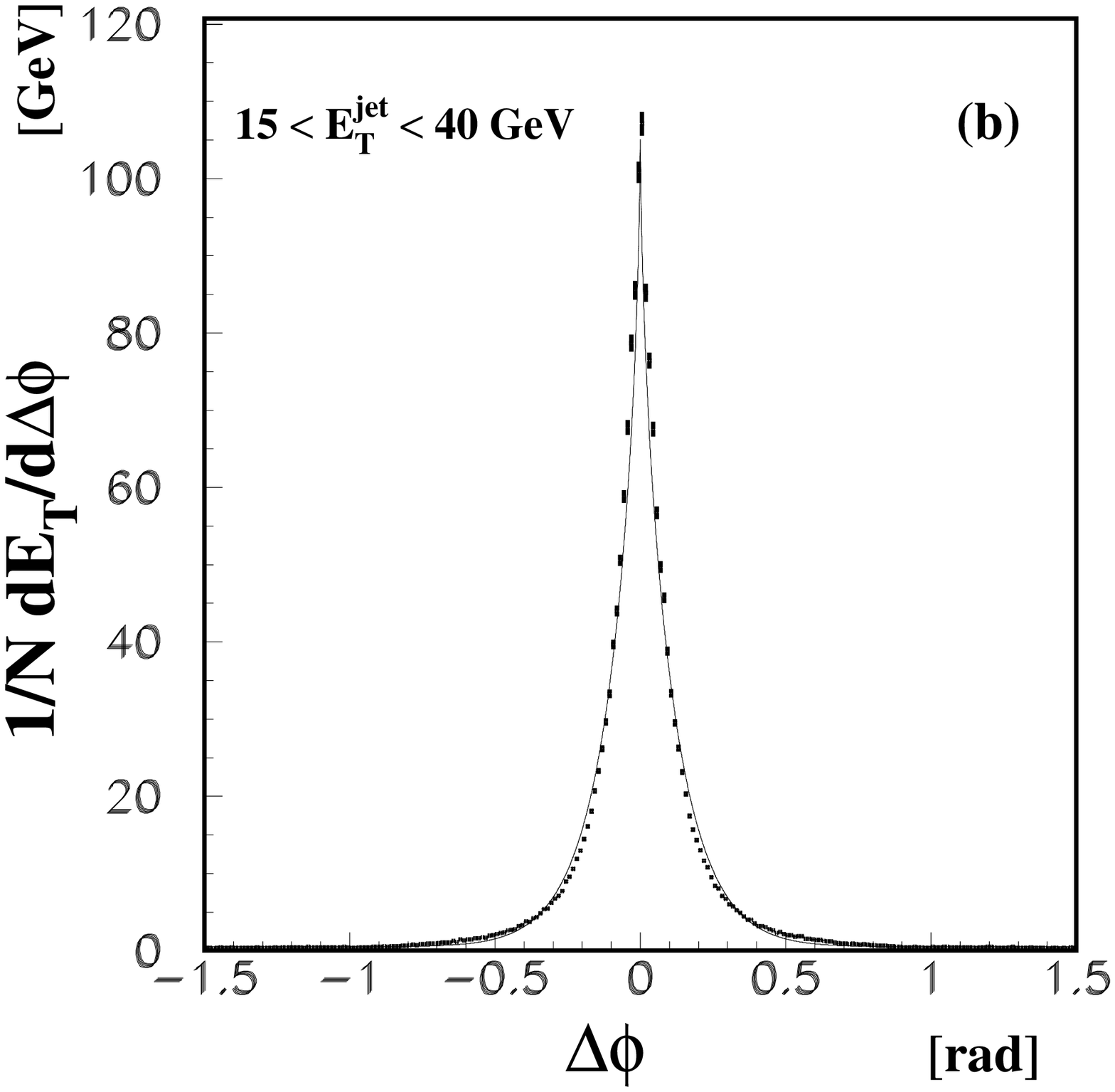}
\end{minipage}
\end{flushleft}
\begin{center}
\caption{
         Transverse energy flows with respect to the jet
         direction projected into the azimuthal direction
         ($\Delta\phi$):
         (a) for the inclusive-jets in $\gamma \gamma$ 
         collisions with $ 3.0 < E_{T}^{jet} < 4.0  {\rm \ GeV}$
         and $|\eta^{jet}|<0.7 $;
         (b) for the inclusive-jets in $e^+e^-$ annihilation
         events with $E_{T}^{jet} > 15  {\rm \ GeV}$.
         Solid line in each figure is the result
         of a fit with a formula given in Eq.(\ref{eq:fit}).}
\label{fig:dEdphi}
\end{center}
\end{figure}

\indent
The transverse energy flow versus the azimuthal angle
around the jet direction is shown in Fig.\ref{fig:dEdphi}a
for inclusive-jets with $3.0 < E_{T}^{jet} < 4.0 \ {\rm GeV}$
and $|\eta^{jet}|<0.7$ in $\gamma\gamma$ collisions.
In a plot the pseudo-rapidity of each
particle is limited to within 1.0 unit
in order to avoid any effect from the remnant-jets.
Fig.\ref{fig:dEdphi}b shows
the same  transverse energy flow for $e^+e^-$ annihilation events with
the same jet cone algorithm. The value
of $E_{T}^{jet}$ is required to be greater than 15 GeV.
Although both energy flows in $\gamma\gamma$ and $e^+e^-$ collisions
show a similar jet profile, the width is very different.

\vspace{5 mm}
In order to make the discussion more quantitative, we define the
jet width as the full width at the half maximum
of the distribution of the transverse
energy flow ($\frac{1}{N}\frac{dE_T}{\Delta\phi}$).
This jet width is
determined by fitting the energy flow with the following formula:
\begin{equation}
f(\Delta \phi) = A \cdot \exp \{ -(\sqrt{|\Delta\phi|}+B)^4
                                + B^4 \} + C,
\label{eq:fit}
\end{equation}
where $\Delta\phi$ is the azimuthal angle difference between
the jet direction and each particle. Parameter $A$ describes the
amplitude of the energy flow at $\Delta\phi = 0$, and $C$ reflects
the constant pedestal energy below the peak.
The jet width ($\mit\Gamma$) can be obtained from the parameter $B$ as
$$
{\mit\Gamma} = 2 \{ (\ln2 + B^4)^{1/4} - B \}^2.
$$
This is the same formula as that used in the
H1 experiment for the analysis of the jet width in $\gamma p$ reactions
\cite{H1gpwidth}.
The fit is carried out only in the peak region, $|\Delta\phi|<1.5$.

\vspace{5 mm}
Since the jet width can be measured by using only a  fiducial volume
of the detector, it is expected that the distortion
caused by the detector is small, but not negligible.
A finite detection efficiency of the charged and neutral particles as
well as their finite resolutions might distort the original jet width.
We correct these detector effects by 
Monte-Carlo simulations. The jet width in the
hadron level $(\mit\Gamma^{cor})$ is obtained in each $E_{T}^{jet}$
bin from the formula
\begin{equation}
       \mit\Gamma^{cor} = \mit\Gamma^{obs} \times
        \frac{1}{\mit\Gamma_{MC}^{obs}/\mit\Gamma_{MC}^{hadron}},
\end{equation}
where $\mit\Gamma^{obs}$ is the width measured in data;
$\mit\Gamma_{MC}^{obs}$ and $\mit\Gamma_{MC}^{hadron}$ are the
width  at the observed and the generator (hadron)
level in the Monte-Carlo simulation, respectively.
The PYTHIA  and PHOJET\cite{PHOJET} programs are
used to determine the correction
factor ($\mit\Gamma_{MC}^{obs}/\mit\Gamma_{MC}^{hadron}$). 
The correction factors ranges
from -15\% to +15\%, depending on the $E_T^{jet}$ region.
The bin-size of  $E_T^{jet}$ is
determined from the consideration of the detector resolution of 
$E_T^{jet}$.

\section{Results and Discussions}
\indent
The results on the jet width ($\mit\Gamma$) for jets produced in
$\gamma\gamma$ collisions
and for $e^+e^-$ annihilation process are summarized 
in Table \ref{tab:jet width}. In this table, the first errors are
statistical and the second ones are systematics.
The systematic errors on the width are estimated by changing
the size of the $\Delta\phi$ bin by factors from 1/5 to 5 to the
nominal bin-size shown in Fig.\ref{fig:dEdphi}.
The systematics caused by these binning effects
are estimated to be less than 3\%. 
The systematics estimated from the difference of the correction
factors in PYTHIA and PHOJET programs are about $\pm$2\%.
In the fit, a pedestal energy below the peak is assumed to be constant
in Eq.(\ref{eq:fit}). The systematics by adding a linear term
are negligible($ \leq 0.1\%$).

\begin{figure}[ht]
\begin{center}
\includegraphics*[height=15cm]{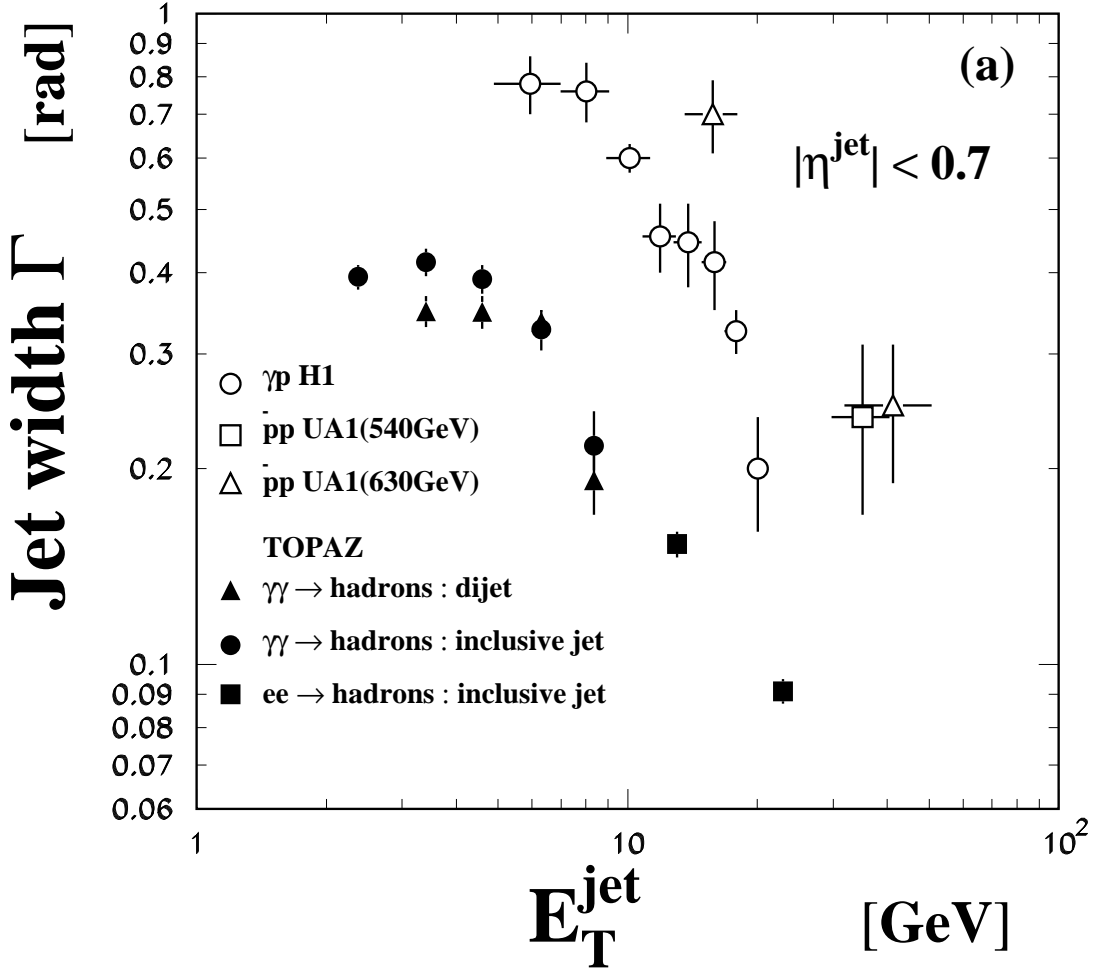}
\caption{
         (a)Jet width as a function of $E_T^{jet}$
         for the inclusive-jets (solid circles) and dijet (solid
         triangles) in $\gamma\gamma$ collisions as well as for the
         inclusive-jets (solid squares)
         in  $e^+e^-$ annihilation process at $\surd{ s_{e^+e^-} }$ = 58
         GeV. The maximum value of $E_{T}^{jet}$ is taken for the dijet
         sample. The error bar shows the statistical error
         and the systematic error added in quadrature.
         For a comparison, the jet widths in other reactions are also
         plotted.
         The open circles show  the jet width in $\gamma p$ collisions
         at $\surd{ s_{ep} }$ = 540 GeV[20]. 
         The open square and open triangles show the width in 
         $ p\bar{p} $ collisions at 
         $\surd{ s }$ = 540 GeV and $\surd{ s }$ = 630 GeV,
         respectively[31].
}
\label{fig:width1_cor}
\end{center}
\end{figure}

\vspace{5 mm}
In Fig.\ref{fig:width1_cor}a, the corrected jet widths
 ($\mit\Gamma$)
in $\gamma\gamma$ interactions as well as the
$e^+e^-$ annihilation events are shown as a function of
$E_{T}^{jet}$. The solid circles are the width from
the inclusive-jets and solid triangles are that from dijet sample.
The solid squares are the jet width for the
$e^+e^-$ annihilation events. 
Since the sample of the inclusive-jets
and dijet events are almost identical in the $e^+e^-$ annihilation events, 
only the results from the inclusive-jets are plotted.
In this figure, the errors 
include statistical and systematic errors added in quadrature.
In the same figure, the results in $\gamma p$ collisions
from H1 experiment at HERA\cite{H1gpwidth}
and $p \bar{p} $ at UA1 experiment\cite{UA1} are also shown
for a comparison.
The following points can be observed from Fig.\ref{fig:width1_cor}a:
\vspace{5 mm}
\begin{enumerate}
\item In the $E_T^{jet}$ region below 4 GeV, 
      the jet width ($\mit\Gamma$) in 
      $\gamma\gamma$ collisions is almost constant.
\item Above $E_T^{jet} >$  4 GeV,
      the width of jets in $\gamma\gamma$ decreases as their 
      $E_T^{jet}$ increases.
\item The jet width in $e^+e^-$ annihilation events located,
      approximately, at the extrapolated points linearly
      on the $E_{T}^{jet}$ dependence in $\gamma\gamma$ collisions.
\item At the same value of $E_{T}^{jet}$, $E_T^{jet}$ = 8 GeV for
      example, the jet width in $\gamma\gamma$ collisions at TRISTAN
      is 3.5-times narrower that that measured in $\gamma p$
      collisions at HERA.
\item The width in  $p \bar{p}$ data from UA1 experiment is
      much wider than that in $\gamma p$ collisions at HERA 
      and also $\gamma\gamma$ collisions at TRISTAN at the same value
      of $E_{T}^{jet}$.
\end{enumerate}
It should be emphasized that the wide dynamic range of  $E_{T}^{jet}$
from 2 GeV to several 10 GeV is covered in  Fig.\ref{fig:width1_cor}a. 

\vspace{5 mm}
According to the suggestions of the QCD analysis of jet shapes 
in refs.\cite{jetwidth1,jetwidth2}, the widths of jets from
$\gamma\gamma$, $\gamma p$ , $p \bar{p}$ 
interactions are plotted as a function of the scaled transverse
energy 
$x_T \equiv 2E_T^{jet}/\sqrt{s}$ in Fig.\ref{fig:width2_cor}b.
Although there are no discussion about
$e^+e^-$ annihilation events in refs.\cite{jetwidth1,jetwidth2},
the jet width in $e^+e^-$ events are also plotted as a function
of $x_{T}$ in the same figure.
In the definition of $x_{T}$, $\sqrt{s}$ represents the
center-of-mass (cms) energy of initial beam particles in
collisions. 
The cms energy of the $e^+e^-$ system
($\sqrt{s_{ee}}=58 {\rm \ GeV}$) is taken
for $\gamma\gamma$ interactions at TRISTAN
\footnote{
  The reason why $\sqrt{s_{ee}}$ should be taken 
  instead of $\sqrt{s_{\gamma\gamma}}$  is 
  caused by the continuous energy spectrum of the photons.
  In the jet cross sections or the jet shape we observe, 
  the photon energy spectrum is integrated over the full
  kinematical regions. As a result, the dependence
  of $\sqrt{s_{\gamma\gamma}}$ disappears and only that of
  $\sqrt{s_{ee}}$  remains.},
 the $ep$ cms energy
($\sqrt{s_{ep}}$ = 297 GeV) is taken for $\gamma p$ interactions
at HERA (H1) and 
$\sqrt{s_{p \bar{p}}}$ = 540 GeV and 630 GeV is taken for UA1
$p \bar{p}$ data.
The $\gamma p$ data at HERA covers 0.02 - 0.1 in  $x_{T}$, 
while the $\gamma\gamma$  data at TRISTAN
covers relatively high-$x_{T}$ regions
of $0.08 - 0.3$.
Two points are plotted for the $e^+e^-$ annihilation events. The lower
point at $x_{T}= 0.42$ corresponds
to the radiative events ($e^+e^- \rightarrow q \bar{q}\gamma$).

One can observe from  Fig.\ref{fig:width2_cor}b 
that the jet width in different reactions and the
different energies are approximately compatible with having
the same dependence on the scaled transverse jet energy $x_{T}$.
It is discussed in refs.\cite{jetwidth1,jetwidth2} that
this kind of scaling behavior is in fact expected from perturbative-QCD
for the jet shape in hadron-hadron and (resolved)$\gamma p$ and
$\gamma\gamma$ collisions. Although the scaling violation is expected 
through
the $E_{T}^{jet}$ dependence in the strong-coupling constant
($\alpha_{s}$), this point is not clear from these data alone.
It is also not clear in this moment
that this kind of argument can be extended
to the $e^+e^-$ annihilation processes\cite{privK}.

\vspace{5 mm}
In order to check the possible difference of the jet width
in direct and resolved processes in $\gamma\gamma$
collisions, the jet width in dijet events was studied.
Fig.\ref{fig:width1_cor_comp}a shows the jet width
for a direct-enriched ($x_{\gamma}^{min} > 0.85$) and
a resolved-enriched ($x_{\gamma}^{min} < 0.85 $) sample. 
The results show that the jet width in the direct sample 
(solid circles) is 
narrower than that of the resolved sample (solid squares) 
by about 0.05 radian 
in the region from $E_{T}^{jet}=$ 4  to 7 GeV.
Above $E_{T}^{jet}>$  7 GeV the statistical 
errors become big.
Since jets originate only from quarks in the direct process
and jets originate from gluons are included in the resolved processes,
this difference might be caused by the shape difference of quark- and
gluon-jets, though the difference is within twice
the standard deviation. 

\vspace{5 mm}
In Fig.\ref{fig:width1_cor_comp}b the jet width in $\gamma\gamma$
collisions and in $e^+e^-$ annihilation events is compared with
PYTHIA predictions. The solid line is the prediction with the default
parameters of the program where
both initial and final parton showers are included. The dotted line is
the one with turning off the parton showers. 
For the hadronization parameters in the program,
the default values which have been determined by 
$ e^+e^-$  annihilation events at LEP1\cite{x} are used.

Fig.\ref{fig:width1_cor_comp}b indicates that in the region 
above 3 GeV the perturbative
parton shower plays an important role concerning the jet width.
The PYTHIA reproduces the measured jet width in $\gamma\gamma$
interactions as well as $e^+e^-$ annihilation events quite well.

\section{Conclusion}
\indent
We have studied the width of the jets produced in $\gamma
\gamma$ collisions as well as $e^+e^-$ annihilation 
process with the jet cone algorithm. 
The jet width in $\gamma\gamma$ collisions becomes narrower
as their $E_T^{jet}$ increases above 4 GeV.
The jet width in $e^+e^-$ annihilation events is found to be 
located at the linearly extrapolated points from the
$E_{T}^{jet}$ dependence in $\gamma\gamma$ collisions. 
Even at the same $E_{T}^{jet}$, a large difference in the jet width
has been observed in jets in $\gamma\gamma$ at TRISTAN,
$\gamma p$ at HERA and $p \bar{p} $ at the UA1 experiment. 
An approximate scaling behavior of the jet width in the different
reaction and the different energy is observed if the width
is plotted as a function of $x_{T}$.

This observation might be important for a comparison of the
jet shape in different reactions.
In ref.\cite{OPALeewidth} the shape difference of the 
quark and gluon are discussed by comparing the shape of the jets in
similar $E_{T}^{jet}$ regions produced in
$e^+e^-$ and $p\bar{p}$\cite{CDFppwidth} collisions. However, the result of
our analysis 
suggests that the jet shape in the different reactions should be
compared at the same $x_{T}$ region.

\section*{Acknowledgment}
\indent
We appreciate useful discussions with Drs. D.~Aurenche, M.~Drees, 
K.~Kramer concerning this analysis.
We thank the TRISTAN accelerator staff
for the successful operation of TRISTAN. We also thank
all of the engineers and technicians at KEK and
the other collaborating institutions: Messrs H.~Inoue, N.~Kimura,
K.~Shiino, M.~Tanaka, K.~Tsukada, N.~Ujiie,
and H.~Yamaoka.

\newpage
\section*{Table caption}
\begin{table}[h]
\newcommand{\lw}[1]{\smash{\lower2.0ex\hbox{#1}}}
\begin{center}
\begin{tabular}{lcccc}
\hline
\lw{Reaction} & $E_T^{jet}$  & $<E_{T}^{jet}>$ & 
        \multicolumn{2}{c}{ Jet Width $(\mit\Gamma)$ } \\
    &\quad (GeV)   & \quad (GeV)    & inclusive-jet &         dijet    \\

\hline\hline
        &
2.0 - 3.0 & 2.37  &0.394 $\pm$ 0.006 $\pm$ 0.016 &                         \\ 
        &
3.0 - 4.0 & 3.40  &0.415 $\pm$ 0.010 $\pm$ 0.017 & 0.349 $\pm$ 0.012 $\pm$ 0.015 \\
$\gamma \gamma$ coll. &
4.0 - 5.5 & 4.59  &0.391 $\pm$ 0.012 $\pm$ 0.016 & 0.348 $\pm$ 0.014 $\pm$ 0.015 \\
        &
5.5 - 7.5 & 6.30  &0.327 $\pm$ 0.018 $\pm$ 0.014 & 0.336 $\pm$ 0.020 $\pm$ 0.014 \\
        &
7.5 - 9.5 & 8.34  &0.217 $\pm$ 0.026 $\pm$ 0.009 & 0.192 $\pm$ 0.021 $\pm$ 0.008 \\
\hline
\lw{$e^+e^-$ annih.} &
10  - 15  & 13.0 & 0.153 $\pm$ 0.003 $\pm$ 0.006 & 0.154 $\pm$ 0.007 $\pm$ 0.006 \\
        &
15  - 40  & 22.9 &0.091 $\pm$ 0.001 $\pm$ 0.004 & 0.087 $\pm$ 0.001 $\pm$ 0.004 \\
\hline
\end{tabular}
\end{center}
\caption{Jet width $(\mit\Gamma)$ in $\gamma\gamma$ collisions and
         $e^+e^-$ annihilation events as a function of $E_T^{jet}$ for
         jets in the central region $|\eta^{jet}|<0.7$.
         $<E_{T}^{jet}>$ is the averaged $E_{T}^{jet}$ weighted by the
         number of events in each $E_{T}^{jet}$ bin. 
         The maximum value of $E_{T}^{jet}$ is taken for a dijet
         sample.
         The first error is statistical and the second error is
         systematic.}
\label{tab:jet width}
\end{table}
%
%
%
%
\newpage
\newpage
\newpage
\newpage
\setcounter{figure}{3}
\newpage
\begin{figure}[ht]
\begin{center}
\includegraphics*[height=15cm]{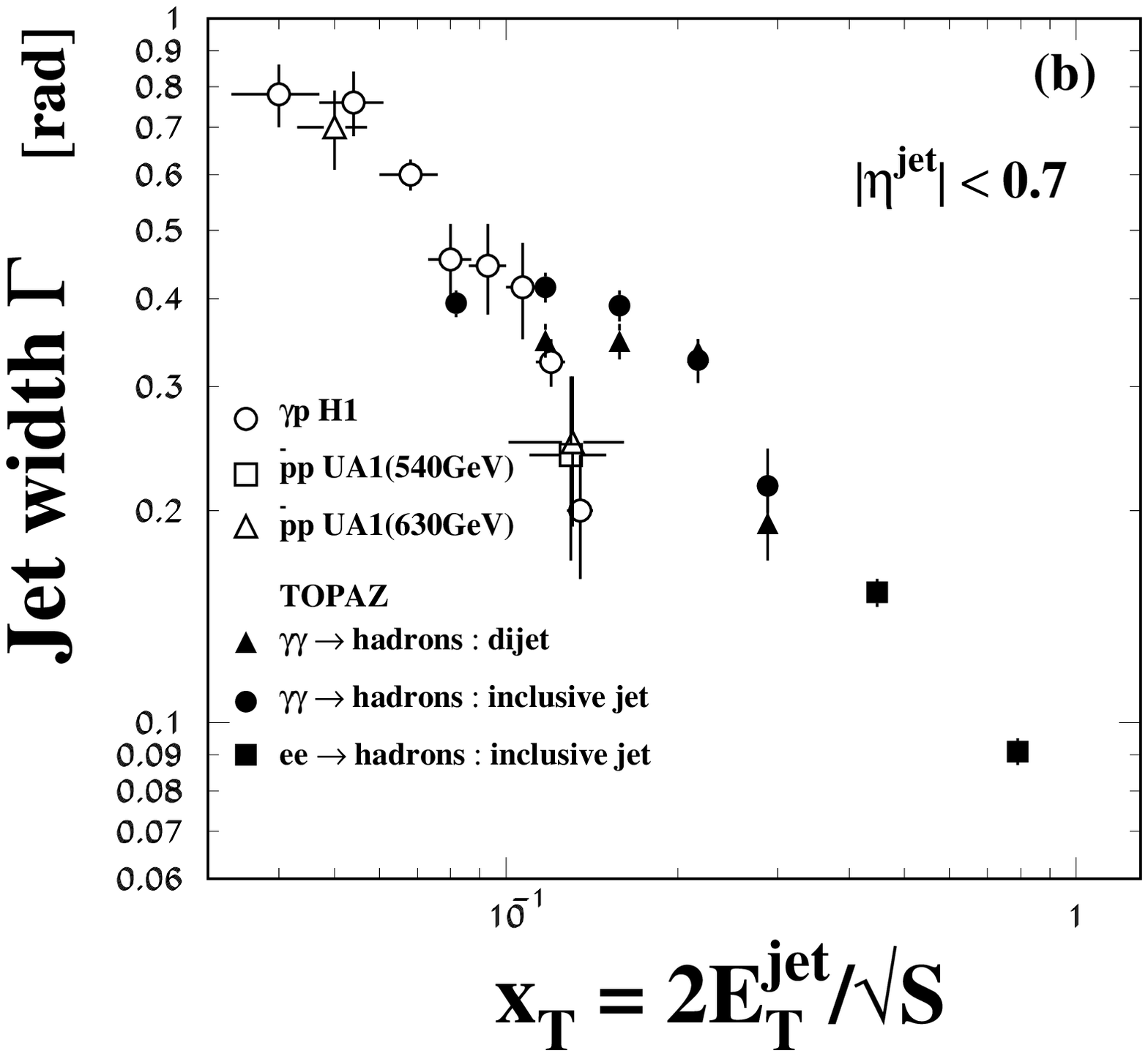}
\caption{
         (b)Jet width as a function of
         $x_T \equiv 2E_T^{jet}/\surd{ s }$, where $\surd{s}$ represents
         the center-of-mass energy of the initial beam system.
         The meaning of the symbols is given in 
         Fig.\ref{fig:width1_cor}a.
         The error bars show the statistical and systematic errors
         added in quadrature.}
\label{fig:width2_cor}
\end{center}
\end{figure}
\begin{figure}[ht]
\begin{flushleft}
\begin{minipage}[htbp]{7.5 cm}
\includegraphics*[height=8.0cm ]{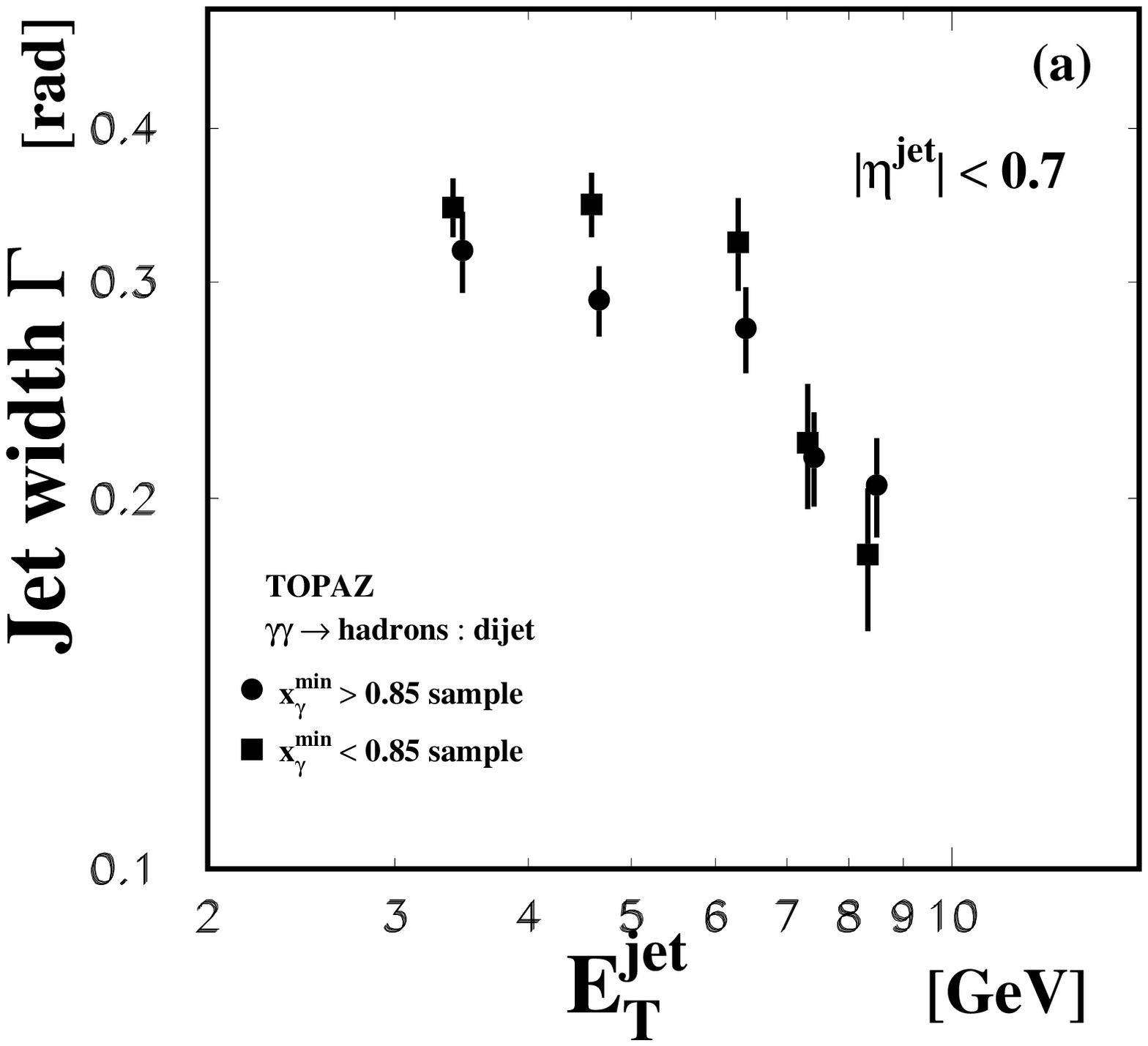}
\end{minipage}
\quad
\begin{minipage}[htbp]{7.5 cm}
\includegraphics*[height=8.0cm]{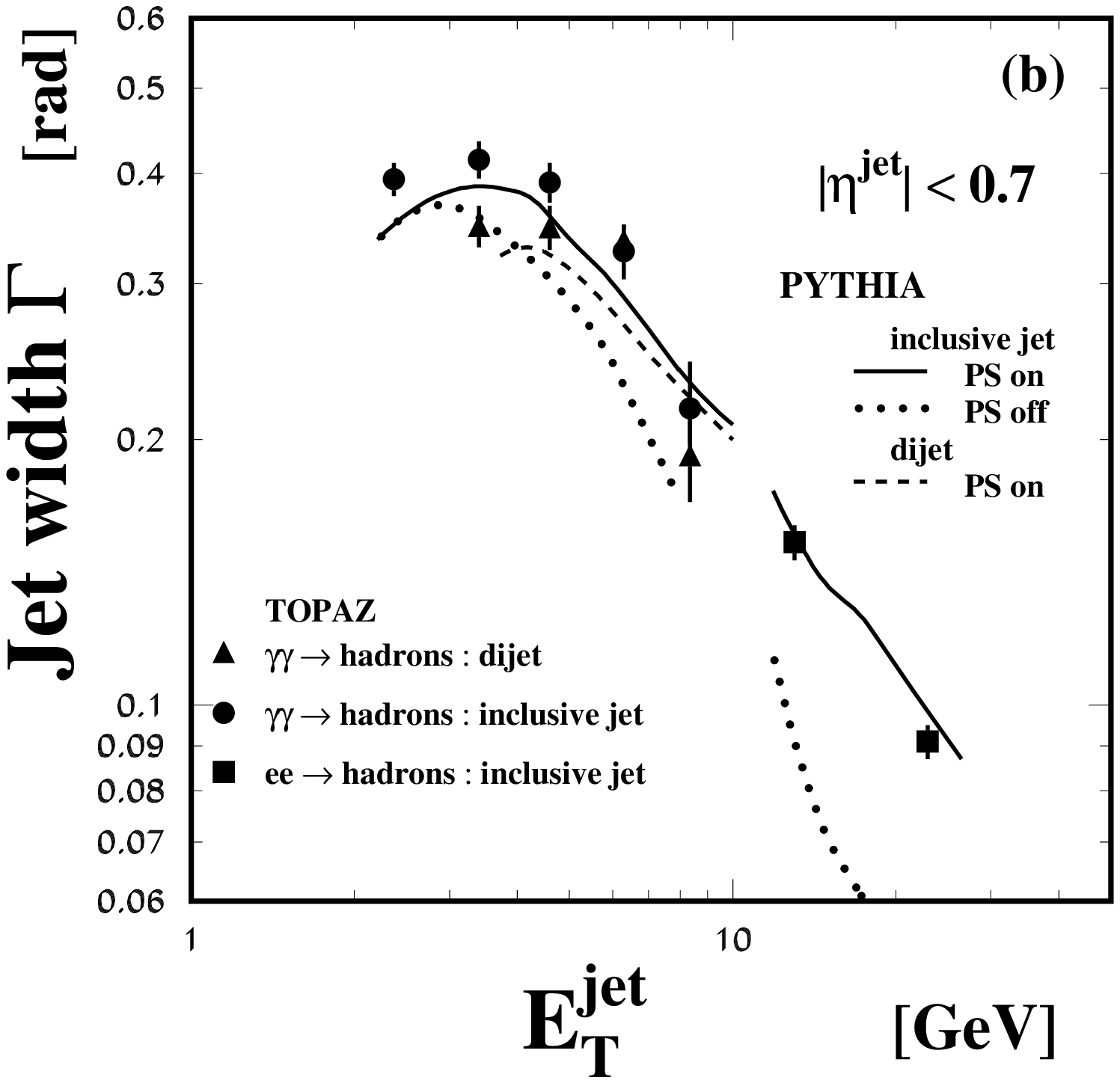}
\end{minipage}
\end{flushleft}
\begin{center}
\caption{
         (a) Comparison of the jet width in direct and resolved
         samples. The solid circles are results for a
         direct sample $( x_{\gamma}^{min} > 0.85$), and the 
         solid squares for a resolved sample ($x_{\gamma} < 0.85$). 
         The error bars show the statistical and systematic errors
         added in quadrature. 
         (b) Comparison of the jet width with PYTHIA Monte-Carlo
         predictions with the parton shower
         (solid line) and without parton shower (dotted line).
         The dashed line shows the Monte-Carlo predictions for a
         dijet sample.}
\label{fig:width1_cor_comp}
\end{center}
\end{figure}

\end{document}